\documentclass[aps,prd,11pt]{revtex4}
\usepackage[colorlinks=true, pdfstartview=FitV, linkcolor=blue, citecolor=red, urlcolor=magenta]{hyperref}
\usepackage{graphicx}
\usepackage{latexsym}
\usepackage{amsmath}
\usepackage{amsfonts}
\usepackage{amssymb}
\usepackage{verbatim}

\usepackage{tikz,tikz-3dplot}
\tdplotsetmaincoords{80}{45}

\tikzset{surface/.style={draw=black, fill=white, fill opacity=.6}}

\usepackage{tikz}
\usetikzlibrary{decorations.pathmorphing,patterns}

\newcommand{\be}{\begin{equation}}
\newcommand{\ee}{\end{equation}}
\newcommand{\bea}{\begin{eqnarray}}
\newcommand{\eea}{\end{eqnarray}}

\newcommand{\ben}{\begin{eqnarray}}
\newcommand{\een}{\end{eqnarray}}

\begin{document}

\title{Holographic complexity of braneworld in Horndeski gravity}

\author{Fabiano F. Santos$^{a}$,  Oleksii Sokoliuk$^{b,c}$ and Alexander Baransky$^{b}$}
\email{fabiano.ffs23-at-gmail.com} \affiliation{$^{a}$Instituto de Física,Universidade Federal do Rio de Janeiro, Caixa Postal 68528, Rio de Janeiro-RJ, 21941-972 -- Brazil}

\email{oleksii.sokoliuk-at-mao.kiev.ua, abaransky-at-ukr.net}\affiliation{$^{b}$Main Astronomical Observatory of the NAS of Ukraine (MAO NASU),
Kyiv, 03143, Ukraine\\
$^{c}$Astronomical Observatory, Taras Shevchenko National University of Kyiv,
3 Observatorna St., 04053 Kyiv, Ukraine}

\begin{abstract}
This work investigates the influence of a probe string on the complexity of braneworld according to the CA (Complexity equals action) conjecture within the Horndeski gravity. In the current study, it is considered that scalar fields that source Horndeski gravity has a spatial dependence. In addition, our system contains a particle moving on the boundary, which corresponds to the insertion of a fundamental string in the higher dimensional bulk. Such an effect is given by the Nambu-Goto term, which also incorporates the time-dependence and evolution in our system. Both warp factor, scalar field, and superpotential values are derived numerically assuming appropriate initial conditions, and the growth rate of holographic complexity is analyzed within the so-called Wheeler-De Witt (WDW) patch with null-like hypersurfaces present.
\end{abstract}
\pacs{xxx,yyy} \maketitle

\section{Introduction}

In recent years, Horndeski's theory of gravitation has been used widely in the studies of black hole information paradox \cite{Santos:2020xox,Santos:2021orr} and holographic thermodynamics (BH entropy, heat capacity, etc.) \cite{Santos:2020xox}. Also, beyond Horndeski theories are now being developed and incorporated \cite{Bravo-Gaete:2020lzs,Sokoliuk:2022llp}. The aforementioned black hole entropy is the thermodynamic quantity, which is extracted through the AdS/BCFT correspondence as presented by the authors \cite{Santos:2021orr,Sokoliuk:2022llp}. This correspondence \cite{Maldacena:1997re,Gubser:1998bc,Witten:1998qj,Aharony:1999ti} is the most important realization of the holographic principle \cite{Susskind:1994vu,tHooft:1993dmi}, which relates a gravity theory in a higher dimensional anti-de Sitter (AdS) bulk to a conformal field theory (CFT), but without gravity living on the bulk boundary. As we know, this theory suggests non-trivial connections between different areas of physics where an example is a particular connection between general relativity and quantum information theory. One of the outstanding developments in this correspondence was proposed by the work of Ryu and Takayanagi \cite{Ryu:2006ef,Ryu:2006bv}, which gives a holographic dictionary for the calculation of entanglement entropy from BCFT. Recently an extension of this proposal was in the Horndeski gravity \cite{Santos:2022oyo}. However, according to the proposal of Ryu and Takayanagi and beyond, the entanglement entropy of the boundary theory is equivalent to the area of a certain minimal surface in the bulk geometry. Thus, we have that the dynamics of the bulk spacetime emerge from the quantum entanglement of the boundary theory \cite{Faulkner:2013ica}. Furthermore, the entanglement entropy may not be enough to probe the sufficient number of degrees of freedom in the black hole interior, since the volume of a black hole usually continues to grow even if spacetime reaches its thermal equilibrium phase \cite{Susskind:2014moa}. It is believed that quantum complexity is the quantity that can continue to grow even after reaching thermal equilibrium, which is similar to the growth of a black hole interior.

\begin{figure}[!htbp]
    \centering
\begin{tikzpicture}[thick,scale=1.8,,
    mycirc/.style={circle,fill=red!70, minimum size=0.01cm}]

 \draw decorate[decoration=zigzag] {(0,0) -- (3,0)};

\draw (0,0) -- (1.5,-1.5);
\draw (3,0) -- (1.5,-1.5);

\draw [ultra thick,fill=blue,draw=orange, fill opacity=0.2, draw opacity = 0.5]
       (3,-1.5) --  (1.5, 0) -- (0, -1.5) -- (1.5,-3) -- cycle;;

\draw decorate[decoration=zigzag] {(0,-3) -- (3,-3)};
\draw[] (3,0) -- (3,-3);

\draw[] (0,0) -- (0,-3);
\node[] at (1.5,0.25) 
    {\large $r=0$};

\node[] at (1.5,-3.25) 
    {\large $r=0$};
    
 \draw (3,0) node[] (C) {};  
        \draw (3,-3) node[] (D) {};  
            \draw[dashed] (C) to [bend left=-15] (D);
 \draw (0,0) node[] (C) {};  
        \draw (0,-3) node[] (D) {};  
            \draw[dashed] (C) to [bend left=15] (D);
\draw (0,-3) -- (1.5,-1.5);
\draw (3,-3) -- (1.5,-1.5);
\node[rotate=-90] at (3.2,-1.5) 
    {\large $r=\infty$};
\node[rotate=90] at (-0.2,-1.5) 
    {\large $r=\infty$};
    
\tikzset{point/.style={insert path={ node[scale=2.5*sqrt(\pgflinewidth)]{.} }}}
\draw[densely dotted, color = black] (1.5,0) -- (1.5,-3) -- cycle;

    \node [fill, draw, circle, minimum width=3pt,color=black, inner sep=0pt,label={[shift={(0.0,0.3)}]\Large$r^{2}_{m}$}] at (1.5,-3) {};
    \node [fill, draw, circle, minimum width=3pt,color=black, inner sep=0pt,label={[shift={(0.0,-1)}]\Large$r^{1}_{m}$}] at (1.5,0) {};   

\node[] at (1.75,-0.8) 
    {\large $I$}; 
\node[] at (2,-1.5) 
    {\large $II$}; 
\node[] at (1.75,-2.2) 
    {\large $III$}; 

\node [fill, draw, circle, minimum width=3pt,color=black, inner sep=0pt,label={[shift={(-0.5,-0.75)}]\Large$r_{\mathrm{max}}$}] at (2.76,-1.25) {};   

\node [fill, draw, circle, minimum width=3pt,color=black, inner sep=0pt] at (2.76,-1.74) {};  

\end{tikzpicture}
\quad
\begin{tikzpicture}[thick,scale=1.8,,
    mycirc/.style={circle,fill=red!70, minimum size=0.01cm}]

 \draw decorate[decoration=zigzag] {(0,0) -- (3,0)};

\draw (0,0) -- (1.5,-1.5);
\draw (3,0) -- (1.5,-1.5);

\draw [ultra thick,fill=blue,draw=orange, fill opacity=0.2, draw opacity = 0.5]
       (3,-1) --  (1.5,-2.5) -- (0,-1) -- (1,0) -- (2,0) -- cycle;

\draw decorate[decoration=zigzag] {(0,-3) -- (3,-3)};
\draw[] (3,0) -- (3,-3);

\draw[] (0,0) -- (0,-3);
\node[] at (1.5,0.25) 
    {\large $r=0$};

\node[] at (1.5,-3.25) 
    {\large $r=0$};
    
 \draw (3,0) node[] (C) {};  
        \draw (3,-3) node[] (D) {};  
            \draw[dashed] (C) to [bend left=-15] (D);
 \draw (0,0) node[] (C) {};  
        \draw (0,-3) node[] (D) {};  
            \draw[dashed] (C) to [bend left=15] (D);
\draw (0,-3) -- (1.5,-1.5);
\draw (3,-3) -- (1.5,-1.5);
\node[rotate=-90] at (3.2,-1.5) 
    {\large $r=\infty$};
\node[rotate=90] at (-0.2,-1.5) 
    {\large $r=\infty$};
    
\node[] at (1.75,-0.8) 
    {\large $I$}; 
\node[] at (2,-1.5) 
    {\large $II$}; 
\node[] at (1.69,-2) 
    {\large $III$}; 

\draw[densely dotted, color = black] (1.5,0) -- (1.5,-2.5) -- cycle;

\node [fill, draw, circle, minimum width=3pt,color=black, inner sep=0pt,label={[shift={(-0.5,0.1)}]\Large$r_{\mathrm{max}}$}] at (2.76,-1.25) {};   

\node [fill, draw, circle, minimum width=3pt,color=black, inner sep=0pt] at (2.82,-0.81) {}; 
\node [fill, draw, circle, minimum width=3pt,color=black, inner sep=0pt,label={[shift={(0.0,-0.7)}]\Large$r^{1}_{m}$}] at (1.5,-2.5) {};   
\end{tikzpicture}
    \caption{Bulk conformal diagram with WDW patch at early ($t_R = t_L = \tau/2 = 0$) and late ($t_R = t_L = \tau/2 > 0$) times with the present singularity at the origin.}
    \label{fig:1.0}
\end{figure}
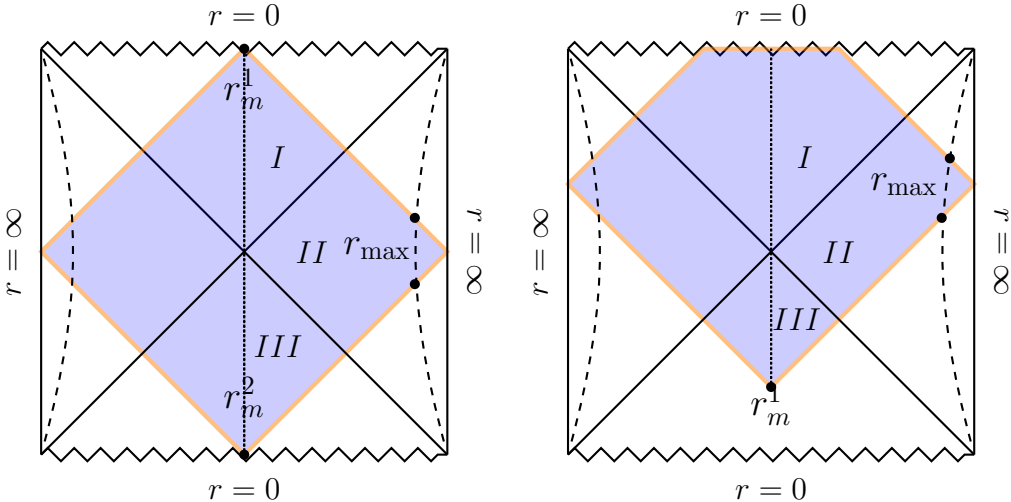

From the viewpoint of quantum information theory, quantum complexity is defined by the minimal number of quantum gates that are needed to build a target state from a reference state \cite{Watrous1,Aaronson:2016vto}.  Additionally, in the framework of AdS/CFT correspondence, one can compute the complexity of states in the boundary quantum field theory of the two-sided AdS black hole through the  $complexity=action$ (CA) conjecture. In this conjecture, it is assumed that quantum complexity on the boundary is associated with the gravitational action evaluated on a region of Wheeler-DeWitt (WDW) patch in the bulk spacetime \cite{Brown:2015bva,Brown:2015lvg}. In the WDW patch, space/time-like boundaries include null boundary surfaces \cite{Hayward1} which can join with each other (for extensive information on the derivation of total WDW action in CA conjecture, refer to the work of \cite{Lehner:2016vdi}). To analyze the evolution of the complexity of braneworld according to the CA conjecture within the Horndeski theory of gravity, we respectively sketch the Penrose diagram of the braneworld causal structure, see Fig.$\sim$\ref{fig:1.0}. In the previously mentioned figure, the WDW patch is denoted by the shaded region, which intersects with cutoffs at times $t_{L}$ and $t_{R}$ respectively. In the present study, we have chosen the symmetrical configuration for the time slices, i.e. $t_{L}=t_{R}\equiv t/2$. Now it will be a handful to evaluate the gravitational action on this patch as the boundary time increases. In our case, the WDW patch includes two UV cutoff surfaces near the asymptotic boundary regions at $r=r_{max}$ which are denoted by black dashed lines on Fig.$\sim$\ref{fig:1.0} and are used to omit IR divergencies. Besides, there are two meeting points in the bulk due to the intersection with the future boundary hypersurface at $r=r^{1}_{m}$ and with the past one at $r=r^{2}_{m}$. As we know, the time evolution from the WDW patch can be encoded in the time dependence of these points.

In this work, we investigate the holographic complexity of the braneworld. This model, being an interesting approach to the hierarchy problem, assumes that the observable universe corresponds to a four-dimensional brane located at the boundary of the ($warped$) extra-dimensional space \cite{Arkani-Hamed:1998jmv,Antoniadis:1998ig,Randall:1999ee,Randall:1999vf,Kaloper:1999sm}. Our boundary term (namely Gibbons-Hawking-Yang term, deformed because of the presence of additional matter fields), does not contribute to the total Nambu-Goto action, since in our configuration we are only taking into account null-like hypersurfaces. Within the braneworld formalism, gravity usually is confined to the brane and therefore holographic confinement could be easily investigated through the Nambu-Goto term with null-like boundaries. The idea behind our prescription is to investigate field derivative coupling.

\section{Methodological route and achievements}

Motivated by the recent applications of the Complexity equals action conjecture we present the investigation of the confinement and complexity of a Horndeski braneworld geometry. Here we present a summary of the main results achieved in this work:

\begin{itemize}

    \item We study the influence of the Horndeski parameters on the braneworld geometry and provide a complete numerical solution for EoMs through first-order formalism; 
    
    \item We construct the braneworld spectrum through the linearization process executing the tensor perturbations around the metric;
    
    \item We construct the null-like boundary term for the Horndeski action integral, which does not contribute to the Nambu-Goto total action; 
   
   \item Additionally, we compute the holographic confinement and complexity of the braneworld (both of the quantities depend on the Horndeski).
\end{itemize}

This work is organized as follows. In Sec.$\sim$\ref{V1}, we present our gravitational setup.  In Sec.$\sim$\ref{V2}, we found the numerical solutions through the first-order formalism. In Sec.$\sim$\ref{V3}, we derive the contribution of the null boundary intersections to the  Gibbons-Hawking term. In Sec.$\sim$\ref{V4}, using the well-known Nambu-Goto action integral, we analyze the holographic confinement with the varying values of modified gravity parameters and we show that the behavior of the energy where $E\sim L$. In Sec.$\sim$\ref{V5}, we present the framework for the holographic complexity and analyze its growth with the variation of free parameters. Finally, in Section \ref{V6} we present our conclusions and final remarks on the key topics of our study.

\section{The Horndeski gravity with a scalar potential}\label{V1}
In the current study, we are going to address the holographic properties of braneworld in the framework of the Horndeski gravity and analyze the linear confinement. The action with a scalar potential within the Horndeski theory is therefore given as follows
\begin{equation}
I[g_{MN},\phi]=\int{\sqrt{-g}d^{5}x\left[\kappa(R-2\Lambda)-\frac{1}{2}(\alpha g_{MN}-\gamma G_{MN})\nabla^{M}\phi\nabla^{N}\phi-V(\phi)\right]}\label{S}
\end{equation}
Note that we have a non-minimal scalar-tensor coupling where we can define a new field $\phi^{'}\equiv\Psi$ with $\kappa=(16\pi G_{N})$, $G_{N}$ being the Newton's gravitational constant. This field has dimension of $(mass)^{2}$ and the parameters $\alpha$ and $\gamma$ control the strength of the kinetic couplings, $\alpha$ is dimensionless and $\gamma$ has dimension of $(mass)^{-2}$. The Einstein field equations can be therefore derived with the use of well-known least-action principle 
\begin{equation}
G_{MN}+\Lambda g_{MN}=\frac{1}{2k}T_{MN}\label{S1}
\end{equation}
where $T_{MN}=\alpha T^{(1)}_{MN}-g_{MN}V(\phi)+\gamma T^{(2)}_{MN}$ with
\begin{equation}\begin{array}{rclrcl}
T^{(1)}_{MN} &=&\nabla_{M}\phi\nabla_{N}\phi-\frac{1}{2}g_{MN}\nabla_{P}\phi\nabla^{P}\phi\\
T^{(2)}_{MN}&=&\frac{1}{2}\nabla_{M}\phi\nabla_{N}\phi R-2\nabla_{P}\phi\nabla_{(M}\phi R^{P}_{N)}-\nabla^{P}\phi\nabla^{K}\phi R_{MPNK}\\
              &-&(\nabla_{M}\nabla^{P}\phi)(\nabla_{N}\nabla_{P}\phi)+(\nabla_{M}\nabla_{N}\phi)\square\phi+\frac{1}{2}G_{MN}(\nabla\phi)^{2}\\
							&-& g_{MN}\left[-\frac{1}{2}(\nabla^{P}\nabla^{K}\phi)(\nabla_{P}\nabla_{K}\phi)+\frac{1}{2}(\square\phi)^{2}-(\nabla_{P}\phi\nabla_{K}\phi)R^{PK}\right]\label{S2}
\end{array}\end{equation}
and the scalar field EoM is 
\begin{equation}
\nabla_{M}[(\alpha g^{MN}-\gamma G^{MN})\nabla_{N}\phi]=V_{\phi}\label{S2}
\end{equation}
The braneworld metric tensor is therefore considered to be a warped one, given by the line element written below
\begin{eqnarray}
ds^{2}&=&g_{MN}dx^Mdx^N\nonumber\\
&=&e^{2A(r)}g_{\mu\nu}dx^{\mu}dx^{\nu}-dr^{2}\label{S4}
\end{eqnarray}
where the latin indices $M,N=$ $0$, $1$, $2$, $3$ and $4$ run on the bulk and the greek indices $\mu,\nu=$ $0$, $1$, $2$ and $3$ run along the braneworld coordinates. 

\section{Equations of motion}\label{v2}
\begin{figure}[!htbp]
    \centering
    \includegraphics[width=\textwidth]{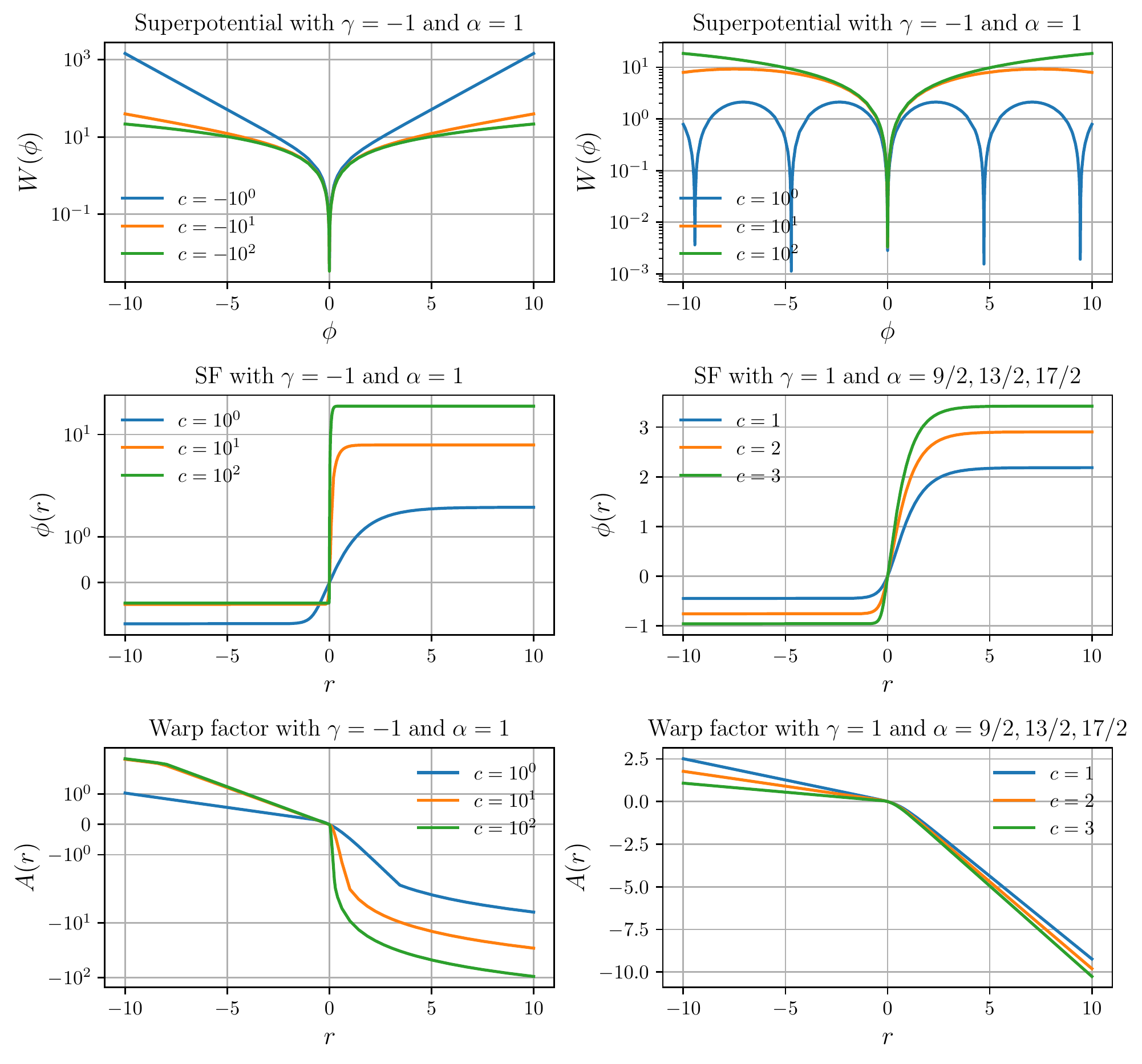}
    \caption{Numerical solutions for superpotential, scalar field and warp factor with varying $c$ and constant $\gamma$, $\alpha$}
    \label{fig:2}
\end{figure}

Now, we present the equations of motions for metric tensor \eqref{S4} combined with the first-order formalism:
\begin{eqnarray}\label{eq:1}
&&A^{'}(r)=-\frac{1}{3}W(\phi)\\
&&\phi^{'}=cW_{\phi}
\end{eqnarray}
For that case, considering the aforementioned EoMs for Horndeski gravitational sector, we have
\begin{eqnarray}
&&2cWW_{\phi\phi}+cW^{2}_{\phi}+\frac{4}{3}W^{2}-\beta=0\label{eq:2}\\
&&\beta=\frac{(2\alpha-b/6c)}{\gamma}\nonumber
\end{eqnarray}
{where $b=24\kappa$. The equation \eqref{eq:2} will be solved numerically using the first order ODE \eqref{eq:1} under the initial conditions $A(0)=\phi(0)=0$, $W(0) =W'(0)=1$. Numerical solutions both provide kink profile and the geometry of braneworld \cite{Brito:2018pwe,Santos:2019ljs,Santos:2021guj}, see Fig.$\sim$\ref{fig:2}. From the aforementioned figure, it is clear that our theoretical prediction has a good agreement with the current state of the Universe. In our case, kink-like solutions behave like the de-Sitter universe in late cosmology, reproducing a dark energy scenario in Horndeski gravity within the first-order formalism due to the scalar field not minimally coupled to the gravitational sector \cite{Santos:2019ljs}. As we can see, the solutions are asymmetrical about the bulk coordinate origin $r=0$, and therefore such a braneworld solution violates the orbifold symmetry $\mathbb{Z}^2$. Furthermore, it was found that $\gamma<1$ increases this asymmetry and that solutions $c<0$ are not physically viable.}

\section{Braneworld spectrum}\label{V2}
For models that consider bosonic fields in a supergravity multiplet, we have six independent wave equations that contribute to glueball states as follows: $J^{PC}=2^{++}$; $1^{+-}$; $1^{--}$; $0^{++}$ and $0^{-+}$ \cite{Brower:2000rp}. In this way, the meson spectrum in AdS/QCD is determined by the solution for eigenmodes of a 5d gauge field living in an AdS slice. As the simple IR cut is very strong, the spectrum of the square of the masses ($m^{2}_{n}$) is similar to that of a Schr\"odinger equation. Thus, in order to analyze such spectrum that is found through the linearization process, we need to execute the tensor perturbations for the metric \eqref{S4} following the steps of \cite{Brito:2018pwe,Fu:2019xtx}. Such methodology, therefore, provides the following:
\begin{eqnarray}
&&C(z)\partial^{2}_{z}h_{\mu\nu}+D(z)\partial_{z}h_{\mu\nu}+\Box_{4d}h_{\mu\nu}=0,\label{T1}\\
&&C(z)=\frac{1-\gamma e^{-2A}\phi^{'2}}{1+\gamma e^{-2A}\phi^{'2}}\label{T2}\\
&&D(z)=\frac{3A^{'}-\gamma e^{-2A}A^{'}\phi^{'2}-2\gamma e^{-2A}\phi^{'2}\phi^{''}}{1+\gamma e^{-2A}\phi^{'2}},\label{T3}
\end{eqnarray}
where through the coordinate transformation $dz=\sqrt{C}d\omega$ it is possible to simplify the equation (\ref{T1}) to:
\begin{eqnarray}
\partial^{2}_{\omega}h_{\mu\nu}+\left(\frac{D}{\sqrt{C}}-\frac{\partial_{\omega}C}{2C}\right)\partial_{\omega}h_{\mu\nu}+\Box_{4d}h_{\mu\nu}=0.\label{T4}
\end{eqnarray} 
where, by using the decomposition $h_{\mu\nu}(x,\omega)=\epsilon_{\mu\nu}(x)e^{-ipx}H(\omega)$ with $p^{2}=-m^{2}$, we have
\begin{eqnarray}
&&\partial^{2}_{\omega}H(\omega)+Q(\omega)\partial_{\omega}H(\omega)+m^{2}H(\omega)=0,\label{T5}\\
&&Q(\omega)=\frac{D}{\sqrt{C}}-\frac{\partial_{\omega}C}{2C}\label{T6},
\end{eqnarray}
However, we can also simplify the equation (\ref{T5}) by redefining $H(\omega)=G(\omega)\psi(\omega)$ with the assumption $G(\omega)=\exp\left(-\frac{1}{2}\int{Q(\omega)d\omega}\right)$. Now, it is possible to derive the so-called Schrödinger-like equation:
\begin{eqnarray}
&&-\partial^{2}_{\omega}\psi(\omega)+U(\omega)\psi(\omega)=m^{2}\psi(\omega),\label{T7}\\
&&U(\omega)=\frac{Q^{2}}{4}+\frac{\partial_{\omega}Q}{2}\label{T8},
\end{eqnarray}
This is an unusual potential as compared with those in the literature. In fact, the potential (\ref{T8}) has the derivative of the scalar field $\phi$. The motivation for the extensive analysis of such potential numerically is that it can provide us with results that generate linear confinement, which is of special interest in our work. Thus, $U(\omega)$ is expected to be Gaussian-like, which in turn leaves the AdS space non-conforming, that is, it implies an energy configuration like $E(L)\sim L$. Furthermore, such an energy configuration is presented in this way, to agree with the linear confinement of $m^{2}_{n}$ growing with $n$, where $L\sim m^{2}_{n} $.

\begin{figure}[!htbp]
    \centering
    \includegraphics[width=0.7\textwidth]{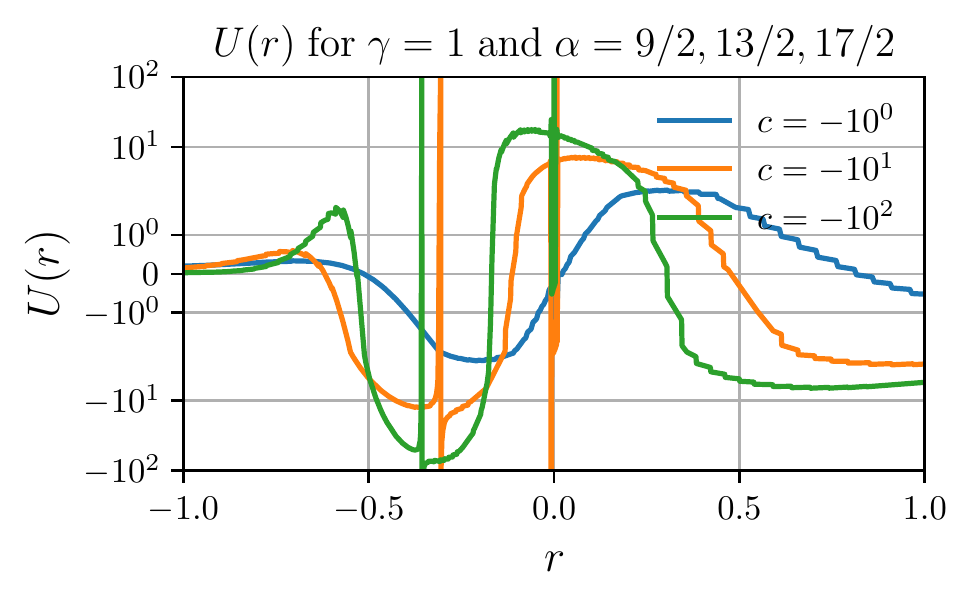}
    \caption{Graviton Kaluza-Klein model wavefunction potential for unitary $\gamma$ and varying positive $c$, $\alpha$}
    \label{fig:3}
\end{figure}

We solve numerically Equation (\ref{T8}) and introduced these solutions graphically in Figure (\ref{fig:3}). As expected, near the braneworld origin (where the brane is usually located), graviton potential behaves as gaussian-like, which ensures the AdS space non-conformality. Such behavior of the graviton potential could only be obtained for $(2\alpha-b/6c)/\gamma\in\mathbb{Z}$ and for positive $\gamma$.

\begin{figure}[!htbp]
    \centering
    \includegraphics[width=\textwidth]{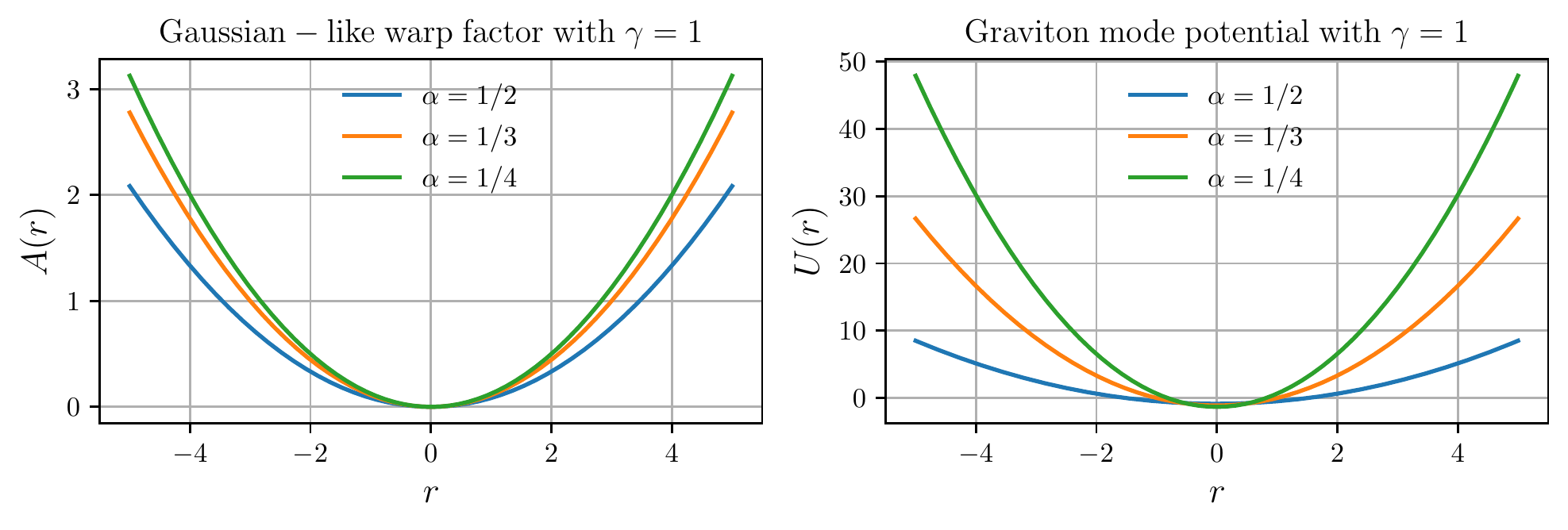}
    \caption{The figure shows the behavior of the gaussian warp factor and potential (\ref{poten2}) for the value of $\gamma=1$ and $\alpha=1/2$ (blue curve), $\alpha=1/3$ (orange curve) and $\alpha=1/4$ (green curve).}
    \label{fig:3.1}
\end{figure}

Because of the fact that four-dimensional gravitational potential has a gaussian-like behavior \cite{Csaki:2000fc,Karch:2000ct}, we can consider a gaussian warp factor for the sake of simplicity
\begin{eqnarray}
A&=&-\frac{1}{2}qr^{2},\label{war}\\
q&=&\frac{(\alpha-1)}{3\gamma}
\end{eqnarray}
where the similar type of geometry has been already considered in confining AdS/QCD \cite{Andreev:2006ct}, for more details regarding the equation (\ref{war}) see \cite{Brito:2018pwe,Fu:2019xtx}. Consequently, we can write the following potential:
\begin{eqnarray}
U(r)=\frac{9q^{2}r^{2}}{2}\frac{(2-\alpha)}{\alpha}-\frac{1}{2}\sqrt{\frac{(2-\alpha)}{\alpha}},\label{poten2}
\end{eqnarray}
In Fig.$\sim$\ref{fig:3.1}, we show the behavior of the aforementioned potential and warp factor for some values of the Horndeski parameters. Judging by the data, extracted from Fig.$\sim$\ref{fig:3.1}, such approximation agrees with the behavior of Fig.$\sim$\ref{fig:3} at some point. 

\section{Null boundaries to the Horndeski gravity}\label{V3}
\begin{figure}[!htbp]
    \centering
\begin{tikzpicture}[thick,scale=1.8,,
    mycirc/.style={circle,fill=red!70, minimum size=0.01cm}]

 \draw decorate[decoration=zigzag] {(0,0) -- (3,0)};

\draw (0,0) -- (1.5,-1.5);
\draw (3,0) -- (1.5,-1.5);

\draw [ultra thick,fill=blue,draw=orange, fill opacity=0.2, draw opacity = 0.5]
       (3,-1.5) --  (1.5, 0) -- (0, -1.5) -- (1.5,-3) -- cycle;;

\draw decorate[decoration=zigzag] {(0,-3) -- (3,-3)};
\draw[] (3,0) -- (3,-3);

\draw[] (0,0) -- (0,-3);
\node[] at (1.5,0.25) 
    {\large $r=0$};

\node[] at (1.5,-3.25) 
    {\large $r=0$};
    
 \draw (3,0) node[] (C) {};  
        \draw (3,-3) node[] (D) {};  
            \draw[dashed] (C) to [bend left=-15] (D);
 \draw (0,0) node[] (C) {};  
        \draw (0,-3) node[] (D) {};  
            \draw[dashed] (C) to [bend left=15] (D);
\draw (0,-3) -- (1.5,-1.5);
\draw (3,-3) -- (1.5,-1.5);
\node[rotate=-90] at (3.2,-1.5) 
    {\large $r=\infty$};
\node[rotate=90] at (-0.2,-1.5) 
    {\large $r=\infty$};
    
\tikzset{point/.style={insert path={ node[scale=2.5*sqrt(\pgflinewidth)]{.} }}}
\draw[densely dotted, color = black] (1.5,0) -- (1.5,-3) -- cycle;

    \node [fill, draw, circle, minimum width=3pt,color=black, inner sep=0pt,label={[shift={(0.0,0.3)}]\Large$r^{2}_{m}$\quad $B_{2}$}] at (1.5,-3) {};
    \node [fill, draw, circle, minimum width=3pt,color=black, inner sep=0pt,label={[shift={(0.0,-1)}]\Large$r^{1}_{m}$\quad $B_{1}$}] at (1.5,0) {};   

\node[] at (1.75,-0.8) 
    {\large $I$}; 
\node[] at (2,-1.5) 
    {\large $II$}; 
\node[] at (1.75,-2.2) 
    {\large $III$}; 

\node [fill, draw, circle, minimum width=3pt,color=black, inner sep=0pt,label={[shift={(-0.5,-0.75)}]\Large$r_{\mathrm{max}}$}] at (2.76,-1.25) {};   

\node [fill, draw, circle, minimum width=3pt,color=black, inner sep=0pt] at (2.76,-1.74) {};  

\end{tikzpicture}

    \caption{Bulk conformal diagram with WDW patch at early ($t_R = t_L = \tau/2 = 0$) and late ($t_R = t_L = \tau/2 > 0$) times with the present singularity at the origin. $r\to 0$ is the singular surface and $r\to\infty$ is the asymptotic boundary surface. The black dashed lines correspond to UV cutoff surfaces at $r=r_{max}$ and $r^{1}_{m}$, $r^{2}_{m}$ are meeting points of null boundaries in the bulk.}
    \label{Null1}
\end{figure}
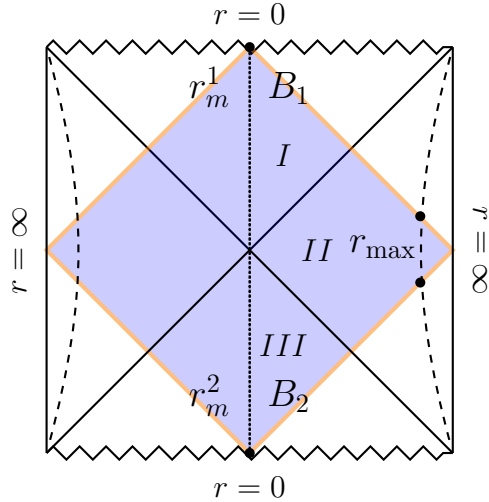

This section introduces the boundary hypersurface action integral. Because of the fact that we have chosen the so-called affine parametrization for the null normals, one could easily notice that the null surface term vanishes. Thus, we only need to consider the Gibbons-Hawking-Yang term coming from the surface at the UV cutoff on the right side of the WDW patch for the Horndeski gravity. Now, it will be a handful to write down the total action:
\begin{eqnarray}
&&I_{total}=\int{\sqrt{-g}d^{5}x\mathcal{L}}-2\kappa\int{d^{4}x\sqrt{\bar{\gamma}}\mathcal{L}_{b}}+2\kappa\int{d^{4}x\sqrt{\bar{\gamma}}\mathcal{L}_{ct}},\label{T1}\\
&&\mathcal{L}=\kappa(R-2\Lambda)-\frac{1}{2}(\alpha g_{MN}-\gamma G_{MN})\nabla^{M}\phi\nabla^{N}\phi\label{T2}\\
&&\mathcal{L}_{b}=K^{({\bar{\gamma}})}-\Sigma^{(\bar{\gamma})}+\frac{\gamma}{4}\left(\nabla_{\mu}\phi\nabla_{\nu}\phi\, n^{\mu}n^{\nu}-(\nabla\phi)^{2}\right)K^{(\bar{\gamma})}+\frac{\gamma}{4}\nabla^{\mu}\phi\nabla^{\nu}\phi K^{(\bar{\gamma})}_{\mu\nu}\label{T3}\\
&&{\cal L}_{ct}=c_{0}+c_{1}R+c_{2}R^{ij}R_{ij}+c_{3}R^{2}+b_{1}(\partial_{i}\phi\partial^{i}\phi)^{2}\label{T4}
\end{eqnarray}
Here $\mathcal{L}_{b}$ corresponds to the Gibbons-Hawking-Yang $\gamma$-dependent term associated with the Horndeski gravity \cite{Santos:2021orr}, where $n^{\mu}$ is an outward pointing unit normal vector to the boundary, $K^{(\bar{\gamma})}=\bar{\gamma}^{\mu\nu}K^{({\bar{\gamma}})}_{\mu\nu}$ is the trace of the extrinsic curvature and $\bar{\gamma}_{\mu\nu}$ is the induced metric on the boundary $r\to\infty$. The Lagrangian ${\cal L}_{ct}$ is related to the boundary counterterms, they do not affect the bulk dynamics and will be neglected. Since we only have null boundaries for the WDW patch, it's more convenient to perform the calculation using the ingoing and outgoing coordinates like:
\begin{eqnarray}
v=t+r^{*}(r);\quad u=t-r^{*}(r),\label{Nu1}
\end{eqnarray}
where $r^{*}$ is a tortoise coordinate defined as
\begin{eqnarray}
r^{*}(r)=\int{e^{-A(r)}dr},\label{Nu2}
\end{eqnarray}
with asymptotic behavior
\begin{eqnarray}
\lim_{r\to\infty}r^{*}(r)=r^{*}_{\infty},\label{Nu3}
\end{eqnarray}
The path includes two UV cutoff surfaces near the asymptotic boundary regions at $r=r_{max}$, denoted by black dashed lines on Fig.$\sim$\ref{fig:1.0}. It is important to mention that null boundaries from the WDW patch are encoded in the time dependence of these points, which satisfy the following relations
\begin{eqnarray}
\frac{t}{2}+r^{*}_{\infty}-r^{*}(r^{1}_{m})=0,\quad \frac{t}{2}-r^{*}_{\infty}+r^{*}(r^{2}_{m})=0\label{Nu3}
\end{eqnarray}
Besides, time evolution for (\ref{Nu2}) is given by the following equalities
\begin{eqnarray}
\frac{dr^{1}_{m}}{dt}=\frac{A(r^{1}_{m})}{2},\quad \frac{dr^{2}_{m}}{dt}=-\frac{A(r^{2}_{m})}{2}\label{Nu4}
\end{eqnarray}
In our prescription, null boundaries of the right sector of the corresponding WDW patch are
\begin{eqnarray}
B_{1}:\frac{t}{2}=r^{*}(r)-r^{*}_{\infty},\quad B_{2}:-\frac{t}{2}=r^{*}(r)-r^{*}_{\infty}\label{Nu5}
\end{eqnarray}
With the above equations, we can study the evolution of the total action (\ref{T1}), see Fig.$\sim$\ref{Null1}. Now, we can define future-directed normal vectors to evaluate $K$ from the equation (\ref{T3})
\begin{eqnarray}
n^{M}=\left(0,0,0,-\frac{z^{'}(r)}{g(r)},\frac{1}{g(r)}\right)\label{Nu6}
\end{eqnarray}
where $g^{2}(r)=1-z^{'2}(r)e^{2A(r)}$ with the induced metric having the form written below
\begin{eqnarray}
ds^{2}_{ind}=e^{2A}(d\tau^{2}+dx^{2}+dy^{2})-g^{2}(r)dr^{2}\label{Nu7}
\end{eqnarray}

Finally, we can derive the exact form of the contribution of surface action in (\ref{T3}):
\begin{eqnarray}
&&I_{bdry}=I^{r=\epsilon}_{bdry}+I^{r=r_{max}}_{bdry}\\
&&=\frac{V_{3}\sqrt{3e^{2A(r)}-g^{2}(r)}}{8\pi G_{N}}\left(\frac{3A^{'}(r)}{g(r)}-\Sigma^{(\bar{\gamma})}+\frac{3\gamma A^{'}\phi^{'2}(r)z^{'2}(r)e^{2A(r)}}{4g^{3}(r)}\right)[(t/2+r^{*}
       _{\infty}-r^{*}(r))|_{r=\epsilon}]\nonumber,
\end{eqnarray}
Since the term above can be removed through the holographic renormalization procedure, it can not contribute to the Nambu-Goto action.
\section{Confinement through the braneworld}\label{V4}

Now we will numerically evaluate whether it is possible to obtain the AdS linearly confining behavior, for which the following holds: $E\sim L$ \cite{Karch:2006pv}. To analyze such behavior, we need to work with the Nambu-Goto action. This well-known action for a probe string can be written as follows
\begin{eqnarray}
S_{NG}=\int{d\sigma d\tau\sqrt{det[\partial_{\alpha}X^{M}\partial_{\beta}X^{N}G_{MN}]}}.\label{NG}
\end{eqnarray}
Through the equation \eqref{NG}, we can construct the corresponding energy associated with this string as
\begin{eqnarray}
E=\pm 2\int^{r_{1}}_{r_{0}}{dr\frac{g(r)f(r)}{\sqrt{f^{2}(r)-f^{2}(r_{0})}}}.\label{NG1}\\
f^{2}(r)=G_{00}G_{x_{||}x_{||}},\quad g^{2}(r)=G_{00}G_{rr}
\end{eqnarray}
For the metric tensor \eqref{S4} the energy and distance are in fact
\begin{eqnarray}
&&E=\pm 2\int^{r_{1}}_{-r_{0}}{dr\frac{e^{A(r)}}{\sqrt{1-e^{4[A(r_{0})-A(r)]}}}},\label{NG2.1}\\
&&L=\pm 2\int^{r_{1}}_{-r_{0}}{dr\frac{e^{2[A(r_{0})-A(r)]}}{\sqrt{1-e^{4[A(r_{0})-A(r)]}}}}\label{NG2.2}
\end{eqnarray}
In order to probe the holographic confinement, one could numerically evaluate the two aforementioned integrals above, which refer to energy and distance. 
\begin{figure}[!htbp]
    \centering
    \includegraphics[width=\textwidth]{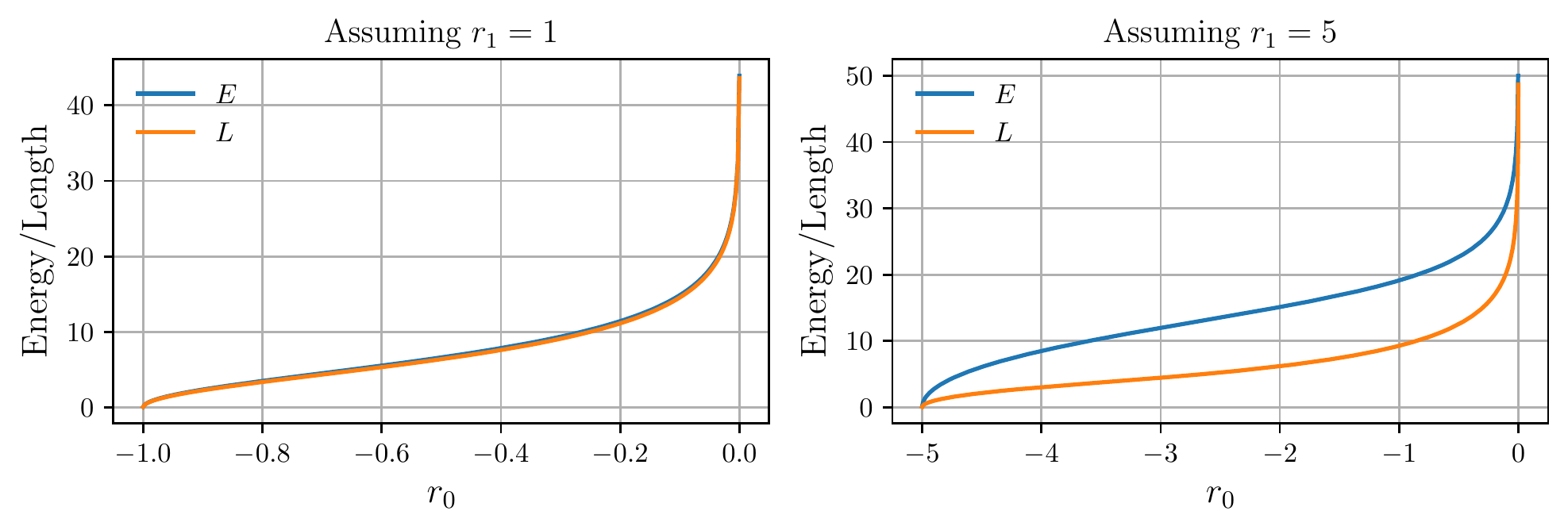}
    \caption{Energy and distance for constant $\gamma=1/2$ and $\alpha=1/4$ and varying $r_0$, $r_1$}
    \label{fig:3.2}
\end{figure}

We plot the numerical solutions for both energy and distance one Fig.$\sim$\ref{fig:3.2}. It could be easily noticed that $E=L$ behavior could be obtained for $\gamma\sim \alpha$ with $r_1\to0$, which leads to the linear confinement of the model. Additional constraints on the values of free parameters $\alpha$ and $\gamma$ are applied by the Gaussian warp factor, since a maximum of the $e^{2A(r)}$ needs to be located at the bulk origin (where brane is located). For some cases, global minimum instead of maximum arises here, which is a non-physically viable behavior of the model. Hence, a model with both viable warp factor form and asymptotically confining with $r_0\to 0$ background could be obtained by applying the following constraints on the parameter space: $|\gamma|\land |\alpha| \ll 1$ and $\gamma<0$.
\section{Evaluating braneworld complexity}\label{V5}
The growth of masses with $n$ in the equation (\ref{T7}), in principle at one point, does not provide us with any explicit example of a string background that behaves in the way we propose. On the other hand, such growth is a generic property of any linear confining gauge theory. Thus, we can formulate and calculate the computational complexity, according to the Complexity equals Action conjecture \cite{Bravo-Gaete:2020lzs,Santos:2020xox}, in order to assess the warp factor influence on the action complexity growth. {Now, following the prescription of \cite{Santos:2022oyo,Feng:2018sqm} and considering the total action(Eq.$\sim$\ref{T1}) with only null boundaries present (therefore, ignoring any joint contributions to the total action integral), we have that the NG action of probe string is given by:}
\begin{eqnarray}
S_{NG}=-T_{s}\int{{\chi}d\sigma\sqrt{-detg_{ind}}},\label{NGA}\\
{\chi=1+2\gamma g^{MN}\nabla_{M}\phi\nabla_{N}\phi}
\end{eqnarray}
To analyze the evolution of the string moving within the warped geometry of a braneworld, we need to construct an induced metric using the parameters $\tau$ and $\sigma$ in the world-sheet of the fundamental string. These parameters are given as follows
\begin{eqnarray}
t=\tau,\quad r=\sigma,\quad\varphi=\omega\tau+\xi(\sigma)\label{NGA1}
\end{eqnarray}
where $\omega$ is a constant angular velocity and $\xi(\sigma)$ is a function that determines the string’s shape. {Thus, we can construct a induced metric for the metric Eq.$\sim$\ref{S4} as}
\begin{eqnarray}
ds^{2}_{ind}=e^{2A(\sigma)}(1+\omega^{2})d\tau^{2}+(e^{2A(\sigma)}\xi^{'2}(\sigma)-1)d\sigma^{2}+2e^{2A(\sigma)}\omega\xi^{'}(\sigma)d\tau d\sigma\label{NGA2}
\end{eqnarray}
Using the above equation, we have
\begin{eqnarray}
&&\frac{dS_{NG}}{T_{s}dt}=\int^{r_{+}}_{r_{-}}{{\chi}d\sigma\mathcal{L}}\label{NGA3}\\
&&\mathcal{L}=e^{A(\sigma)}\sqrt{-1-\omega^{2}+\xi^{'2}J(\sigma)}\label{NGA4}\\
&&J(\sigma)=e^{2A(\sigma)}+\omega^{2}(1+e^{2A(\sigma)})\nonumber
\end{eqnarray}
The equations of motion to the Lagrangian (\ref{NGA4}) are given by
\begin{eqnarray}
\frac{d}{d\sigma}\left(\frac{\partial\mathcal{L}}{\partial\xi^{'}(\sigma)}\right)-\frac{\partial\mathcal{L}}{\partial\xi(\sigma)}=0.\label{NGA5}
\end{eqnarray}
An solving the above equation, we can find
\begin{eqnarray}
\xi^{'}=c_{\xi}\sqrt{\frac{1-\omega^{2}}{J^{2}e^{2A(\sigma)}-c^{2}_{\xi}J}}\label{NGA6}
\end{eqnarray}
Choosing $c_{\xi}=1$ the Lagrangian (\ref{NGA4}) becomes
\begin{eqnarray}
\mathcal{L}=e^{A(\sigma)}\sqrt{\frac{2-(1+\omega^{2})Je^{2A(\sigma)}}{Je^{2A(\sigma)}-1}}.\label{NGA7}
\end{eqnarray}
Now, rewrite the equation (\ref{NGA3}) with the equation (\ref{NGA7}), we can conclude that
\begin{eqnarray}
&&\frac{dS_{NG}}{T_{s}dt}=\int^{r_{+}}_{r_{-}}{{\chi}d\sigma e^{A(\sigma)}\sqrt{\frac{2-(1+\omega^{2})Je^{2A(\sigma)}}{Je^{2A(\sigma)}-1}}}\label{NGA8}\\
&&A=-\frac{1}{2}q\sigma^{2}\label{NGA9}\\
&&{\chi=1-6\gamma cA^{''}=1+2c(\alpha-1)}
\end{eqnarray}
{Here, we are going to use $\chi\sim 1$, also assuming that $c=1/24$ and $\alpha=1/4$}. The equation (\ref{NGA8}) must be analyzed numerically where the analytical solution to the superpotential with gaussian-like warp-factor comes from the equation (\ref{eq:2}). Remarkably, even for the warp factor that possesses orbifold symmetry, solutions for the scalar field are not symmetrical w.r.t. bulk coordinate origin and superpotential have highly oscillatory behavior.

\begin{figure}[!htbp]
    \centering
    \includegraphics[width=\textwidth]{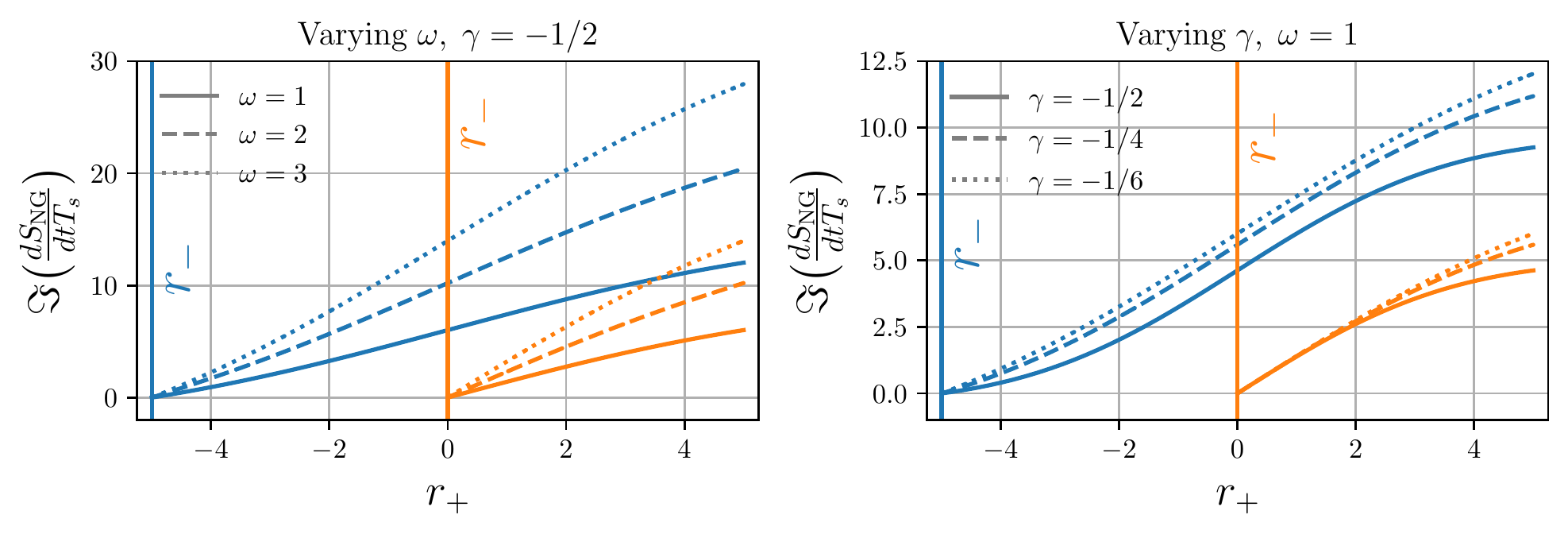}
    \caption{Imaginary part of holographic complexity temporal derivative numerical behavior normalized by the string tension $T_s$ with varying values of string angular momentum $\omega$ and Horndeski parameter $\gamma$ (we assume that $c=1/24$ and $\alpha=1/4$). We have chosen such values of free parameters that AdS background is linearly confining one}
    \label{fig:last}
\end{figure}
We plot the numerical behavior of the Numbu-Goto string holographic complexity temporal derivative routinely on the Figure (\ref{fig:last}) for different $r_-$. As we noticed during the numerical analysis, for linearly confining background and physically viable form of Gaussian-like warp factor, $\frac{dS_{\mathrm{NG}}}{dt T_s}$ is imaginary, so that $\Re(\frac{dS_{\mathrm{NG}}}{dt T_s})=0$. Moreover, values of this quantity generally grow with $\omega\to\infty$, which was also observed for AdS$_5$ BH solution with rotating string ($\omega\neq0$) background (for more information on the subject, see \cite{Nagasaki:2017kqe}). Besides, a growing trend could also be observed for limit $\gamma\to0$. Also, as expected, with the bigger area (i.e. spacing between $r_-$ and $r_+$) normalized temporal derivative values of holographic complexity grow non-linearly since it will be harder to retrieve the initial state with a bigger area of computation, and therefore $S_{\mathrm{NG}}$ will grow faster. {In this case, our results lead to the inclusion of Coulomb contributions which accelerate the glueball particles to the desired velocity. In that case, the energy of the glueball can be obviously defined as its initial rest energy plus the work done by the electric field while accelerating the glueball. Acceleration time from such contributions should be small compared to the damping time to avoid counting energy that has already been lost to the medium. On the other hand, the acceleration time should also be large compared to the inverse kinetic energy of the glueball, to minimize the uncertainty of the energy estimation and localize the initial state of the brane with more precision}.
\section{Conclusions}\label{V6}
In the current paper, we comprehensively investigated the physics around Horndeski-Randall-Sundrum braneworlds (endangered by the non-minimally coupled to gravity scalar field) using various numerical methods and assuming different scenarios with varying values of modified gravitation-free parameters, such as $\alpha$ and $\gamma$ that define the strength of interactions between scalar field and gravity.

At first, we derived equations of motion for classic matter content and scalar fields using the well-known least-action principle $\delta S=0$. By imposing initial conditions on the scale factor, scalar field, and superpotential we found their numerical forms and respectively plotted them in the Figure (\ref{fig:2}).

Furthermore, we carefully investigated the braneworld spectrum, for tensor modes $h_{\mu\nu}$ we found corresponding Schr\"odinger-like equation and potential that in turn appears to be Gaussian-like (see Figure (\ref{fig:3})), which could lead to the non-conforming AdS spacetime. Because of such Gaussian-like behavior, we considered the Gaussian-like warp factor for the sake of simplicity, solutions for which are plotted in Figure (\ref{fig:3.1}).
               
Besides the aforementioned studies, we as well probed the linear confinement of our Horndeski braneworld model with the present Nambu-Goto string. From the numerical evaluation of both (\ref{NG2.1}) and (\ref{NG2.2}) integrals, we obtained the values of free parameters, for which asymptotically $E=L$, which is required for linear confinement (for graphical representation, see Figure (\ref{fig:3.2})).

Finally, we wanted to study the behavior of holographic complexity and its temporal derivative for our linearly confining Horndeski braneworld. With the use of Nambu-Goto probe string action and induced on the string metric tensor (we assume that our NG string is rotating, i.e. generally angular velocity $\omega$ is constant but non-vanishing). We derived normalized by string tension temporal derivative of NG action in the Equation (\ref{NGA8}) and found numerical solutions for the aforementioned quantity $\frac{dS_{\mathrm{NG}}}{T_sdt}$ with linearly confining background spacetime, placed those solutions on the Figure (\ref{fig:last}). As it was unveiled, this quantity is completely imaginary, so the real counterpart does vanish. Values of those solutions do grow with $\omega\to\infty$. On the other hand, they grow as well with $\gamma\to0$, so coupling the scalar field to gravity reduces the complexity growth rate. Such behavior as it was already stated could lead to the inclusion of Coulomb contributions {to minimize the uncertainty of the energy estimation and localize the initial state of the brane with more precision}.

\acknowledgments
Sokoliuk O. performed the work in the frame of the "Mathematical modeling in interdisciplinary research of processes and systems based on intelligent supercomputer, grid and cloud technologies" program of the NAS of Ukraine. We are very much grateful to the honorable referees and to the editor for the illuminating suggestions that have significantly improved our work in terms of research quality, and presentation.


\begin{thebibliography}{99}

\bibitem{Santos:2020xox}
F.~F.~Santos,
{\it Rotating black hole with a probe string in Horndeski Gravity},
Eur. Phys. J. Plus \textbf{135}, no.10, 810 (2020),
[arXiv:2005.10983 [hep-th]].


\bibitem{Santos:2021orr}
F.~F.~Santos, E.~F.~Capossoli and H.~Boschi-Filho,
{\it AdS/BCFT correspondence and BTZ black hole thermodynamics within Horndeski gravity},
Phys. Rev. D \textbf{104}, no.6, 066014 (2021),
[arXiv:2105.03802 [hep-th]].

\bibitem{Bravo-Gaete:2020lzs}
M.~Bravo-Gaete and F.~F.~Santos,
{\it Complexity of four-dimensional hairy anti-de-Sitter black holes with a rotating string and shear viscosity in generalized scalar\textendash{}tensor theories},
Eur. Phys. J. C \textbf{82}, no.2, 101 (2022)
doi:10.1140/epjc/s10052-022-10064-y
[arXiv:2010.10942 [hep-th]].

\bibitem{Sokoliuk:2022llp}
O.~Sokoliuk, F.~F.~Santos and A.~Baransky,
{\it AdS/BCFT correspondence and Lovelock theory in the presence of canonical scalar field},
[arXiv:2206.04054 [hep-th]].

  \bibitem{Maldacena:1997re} 
  J.~M.~Maldacena,
  {\it The Large N limit of superconformal field theories and supergravity},
  Int.\ J.\ Theor.\ Phys.\  {\bf 38}, 1113 (1999)
  [Adv.\ Theor.\ Math.\ Phys.\  {\bf 2}, 231 (1998)]
    [hep-th/9711200].
	
  \bibitem{Gubser:1998bc} 
  S.~S.~Gubser, I.~R.~Klebanov and A.~M.~Polyakov,
  {\it Gauge theory correlators from noncritical string theory},
  Phys.\ Lett.\ B {\bf 428}, 105 (1998),
  [hep-th/9802109].
  
  \bibitem{Witten:1998qj} 
  E.~Witten,
  {\it Anti-de Sitter space and holography},
  Adv.\ Theor.\ Math.\ Phys.\  {\bf 2}, 253 (1998),
  [hep-th/9802150].
	
  \bibitem{Aharony:1999ti} 
  O.~Aharony, S.~S.~Gubser, J.~M.~Maldacena, H.~Ooguri and Y.~Oz,
  {\it Large N field theories, string theory and gravity},
  Phys.\ Rept.\  {\bf 323}, 183 (2000),
  [hep-th/9905111].

\bibitem{Susskind:1994vu}
L.~Susskind,
{\it The World as a hologram},
J. Math. Phys. \textbf{36}, 6377-6396 (1995)
doi:10.1063/1.531249
[arXiv:hep-th/9409089 [hep-th]].

\bibitem{tHooft:1993dmi}
G.~'t Hooft,
{\it Dimensional reduction in quantum gravity},
Conf. Proc. C \textbf{930308}, 284-296 (1993)
[arXiv:gr-qc/9310026 [gr-qc]].

\bibitem{Ryu:2006ef}
S.~Ryu and T.~Takayanagi,
{\it Aspects of Holographic Entanglement Entropy},
JHEP \textbf{08} (2006), 045,
[arXiv:hep-th/0605073 [hep-th]].

\bibitem{Ryu:2006bv}
S.~Ryu and T.~Takayanagi,
{\it Holographic derivation of entanglement entropy from AdS/CFT},
Phys. Rev. Lett. \textbf{96} (2006), 181602,
[arXiv:hep-th/0603001 [hep-th]].

\bibitem{Santos:2022oyo}
F.~F.~Santos,
{\it Entanglement Entropy in Horndeski Gravity},
doi:10.22128/jhap.2022.488.1015
[arXiv:2201.02500 [hep-th]].



\bibitem{Faulkner:2013ica}
T.~Faulkner, M.~Guica, T.~Hartman, R.~C.~Myers and M.~Van Raamsdonk,
{\it Gravitation from Entanglement in Holographic CFTs},
JHEP \textbf{03}, 051 (2014)
doi:10.1007/JHEP03(2014)051
[arXiv:1312.7856 [hep-th]].

\bibitem{Susskind:2014moa}
L.~Susskind,
{\it Entanglement is not enough},
Fortsch. Phys. \textbf{64}, 49-71 (2016)
doi:10.1002/prop.201500095
[arXiv:1411.0690 [hep-th]].

\bibitem{Watrous1}
J. Watrous, 
{\it Quantum computational complexity," Encyclopedia of Complexity and Systems Science, ed.,
R. A. Meyers}
7174-7201 (2009), [arXiv:0804.3401 [quant-ph]]

\bibitem{Aaronson:2016vto}
S.~Aaronson,
{\it The Complexity of Quantum States and Transformations: From Quantum Money to Black Holes},
[arXiv:1607.05256 [quant-ph]].

\bibitem{Brown:2015bva}
A.~R.~Brown, D.~A.~Roberts, L.~Susskind, B.~Swingle and Y.~Zhao,
{\it Holographic Complexity Equals Bulk Action?},
Phys. Rev. Lett. \textbf{116}, no.19, 191301 (2016)
doi:10.1103/PhysRevLett.116.191301
[arXiv:1509.07876 [hep-th]].

\bibitem{Brown:2015lvg}
A.~R.~Brown, D.~A.~Roberts, L.~Susskind, B.~Swingle and Y.~Zhao,
{\it Complexity, action, and black holes},
Phys. Rev. D \textbf{93}, no.8, 086006 (2016)
doi:10.1103/PhysRevD.93.086006
[arXiv:1512.04993 [hep-th]].

\bibitem{Hayward1}
G. Hayward, {\it Gravitational action for space-times with nonsmooth boundaries}, Phys. Rev. D 47 (1993),
3275-3280.

\bibitem{Lehner:2016vdi}
L.~Lehner, R.~C.~Myers, E.~Poisson and R.~D.~Sorkin,
{\it Gravitational action with null boundaries},
Phys. Rev. D \textbf{94}, no.8, 084046 (2016)
doi:10.1103/PhysRevD.94.084046
[arXiv:1609.00207 [hep-th]].

\bibitem{Arkani-Hamed:1998jmv}
N.~Arkani-Hamed, S.~Dimopoulos and G.~R.~Dvali,
{\it The Hierarchy problem and new dimensions at a millimeter},
Phys. Lett. B \textbf{429}, 263-272 (1998)
doi:10.1016/S0370-2693(98)00466-3
[arXiv:hep-ph/9803315 [hep-ph]].

\bibitem{Antoniadis:1998ig}
I.~Antoniadis, N.~Arkani-Hamed, S.~Dimopoulos and G.~R.~Dvali,
{\it New dimensions at a millimeter to a Fermi and superstrings at a TeV},
Phys. Lett. B \textbf{436}, 257-263 (1998)
doi:10.1016/S0370-2693(98)00860-0
[arXiv:hep-ph/9804398 [hep-ph]].

\bibitem{Randall:1999ee}
L.~Randall and R.~Sundrum,Kaloper:1999sm
{\it A Large mass hierarchy from a small extra dimension},
Phys. Rev. Lett. \textbf{83}, 3370-3373 (1999)
doi:10.1103/PhysRevLett.83.3370
[arXiv:hep-ph/9905221 [hep-ph]].

\bibitem{Randall:1999vf}
L.~Randall and R.~Sundrum,
{\it An Alternative to compactification},
Phys. Rev. Lett. \textbf{83}, 4690-4693 (1999)
doi:10.1103/PhysRevLett.83.4690
[arXiv:hep-th/9906064 [hep-th]].

\bibitem{Kaloper:1999sm}
N.~Kaloper,
{\it Bent domain walls as brane worlds},
Phys. Rev. D \textbf{60}, 123506 (1999)
doi:10.1103/PhysRevD.60.123506
[arXiv:hep-th/9905210 [hep-th]].


\bibitem{Brito:2018pwe}
F.~A.~Brito and F.~F.~Dos Santos,
{\it Braneworlds in Horndeski gravity},
[arXiv:1810.08196 [hep-th]].

\bibitem{Santos:2019ljs}
F.~F.~Santos, R.~M.~P.~Neves and F.~A.~Brito,
{\it Modeling dark sector in Horndeski gravity at first-order formalism},
Adv. High Energy Phys. \textbf{2019}, 3486805 (2019),
[arXiv:1906.11821 [hep-th]].

\bibitem{Santos:2021guj}
F.~F.~Santos and F.~A.~Brito,
{\it Domain walls in Horndeski gravity},
[arXiv:2105.00343 [hep-th]].


\bibitem{Brower:2000rp}
R.~C.~Brower, S.~D.~Mathur and C.~I.~Tan,
{\it Glueball spectrum for QCD from AdS supergravity duality},
Nucl. Phys. B \textbf{587}, 249-276 (2000),
[arXiv:hep-th/0003115 [hep-th]].

\bibitem{Sui:2021uic}
T.~T.~Sui, Y.~P.~Zhang, B.~M.~Gu and Y.~X.~Liu,
Eur. Phys. J. C \textbf{81} (2021) no.11, 980
doi:10.1140/epjc/s10052-021-09756-8

\bibitem{Fu:2019xtx}
Q.~M.~Fu, H.~Yu, L.~Zhao and Y.~X.~Liu,
{\it Thick brane in reduced Horndeski theory},
Phys. Rev. D \textbf{100}, no.12, 124057 (2019)
doi:10.1103/PhysRevD.100.124057
[arXiv:1907.12049 [gr-qc]].

\bibitem{Karch:2006pv}
A.~Karch, E.~Katz, D.~T.~Son and M.~A.~Stephanov,
{\it Linear confinement and AdS/QCD},
Phys. Rev. D \textbf{74} (2006), 015005
doi:10.1103/PhysRevD.74.015005
[arXiv:hep-ph/0602229 [hep-ph]].

\bibitem{Csaki:2000fc}
C.~Csaki, J.~Erlich, T.~J.~Hollowood and Y.~Shirman,
{\it Universal aspects of gravity localized on thick branes},
Nucl. Phys. B \textbf{581}, 309-338 (2000),
[arXiv:hep-th/0001033 [hep-th]].

\bibitem{Karch:2000ct}
A.~Karch and L.~Randall,
{\it Locally localized gravity},
JHEP \textbf{05}, 008 (2001),
[arXiv:hep-th/0011156 [hep-th]].
\bibitem{Andreev:2006ct} 
  O.~Andreev and V.~I.~Zakharov,
  {\it Heavy-quark potentials and AdS/QCD},
  Phys.\ Rev.\ D {\bf 74}, 025023 (2006),
  [hep-ph/0604204].


\bibitem{Feng:2018sqm}
X.~H.~Feng and H.~S.~Liu,
{\it Holographic Complexity Growth Rate in Horndeski Theory},
Eur. Phys. J. C \textbf{79}, no.1, 40 (2019),
[arXiv:1811.03303 [hep-th]].

\bibitem{Nagasaki:2017kqe}
K.~Nagasaki,
{\it Complexity of AdS$_5$ black holes with a rotating string},
Phys. Rev. D \textbf{96}, no.12, 126018 (2017)
doi:10.1103/PhysRevD.96.126018
[arXiv:1707.08376 [hep-th]].

\end{thebibliography}
\end{document}